\documentclass[11pt, oribibl]{llncs}
\usepackage{graphicx}
\usepackage{multirow}
\begin{document}
\frontmatter          
\pagestyle{headings}  
\addtocmark{Cache Oblivious Priority Queues and Shortest Path Problems} 

\mainmatter              
\title{An Empirical Study of Cache-Oblivious Priority Queues and their Application to the Shortest Path Problem}
\titlerunning{Empirical CO-SSSP on Massive Graphs}  
%
\author{Benjamin Sach and Rapha\"{e}l Clifford}
\authorrunning{Sach and Clifford}   
%
\tocauthor{Benjamin Sach,Rapha\"{e}l Clifford}
\institute{Bristol University, Bristol, UK\\
\email{\{sach,clifford\}@cs.bris.ac.uk},\\
\texttt{http://www.cs.bris.ac.uk/$\sim$sach/COSP/}}

\maketitle              

\begin{abstract}
In recent years the Cache-Oblivious model of external memory computation has provided an attractive theoretical basis for the analysis of algorithms on massive datasets.  Much progress has been made in discovering algorithms that are asymptotically optimal or near optimal.  However, to date there are still relatively few successful experimental studies. In this paper we compare two different Cache-Oblivious priority queues based on the Funnel  and Bucket Heap and apply them to the single source shortest path problem on graphs with positive edge weights.   Our results show that  when RAM is limited and data is swapping to external storage, the Cache-Oblivious priority queues achieve orders of magnitude speedups over standard internal memory techniques.  However, for the single source shortest path problem both on simulated and real world graph data, these speedups are markedly lower due to the time required to access the graph adjacency list itself.     
\end{abstract}

%

\section{Introduction} 


The need to transfer blocks of data between memory levels is a property of real world systems not accounted for in the standard RAM model of computing. The I/O-Model introduced by Aggarwal and Vitter~\cite{Aggarwal}, considers two levels of memory, internal and external. The internal memory is of fixed size $M$ and the external memory is unbounded in size. Data is transferred between levels in blocks of size $B$ with each block transferred costing a single I/O operation. 


Frigo et al~\cite{Frigo} later introduced the Cache-Oblivious model which provides a theoretical basis for designing algorithms for systems with multiple levels of memory. This model has two significant advantages. First, algorithms designed specifically for the standard two level I/O model (so-called Cache-Aware algorithms) need careful tuning to the parameters of the system on which they are run. More significantly, modern computer systems may contain many levels of cache, internal memory and external storage. An optimal Cache-Oblivious algorithm will in theory be optimal across all levels of the memory hierarchy simultaneously~\cite{Frigo}. 

There has been a flurry of results in Cache-Oblivious algorithms since its conception which include sorting, linked lists, B-trees, orthogonal range searching and priority queues (see e.g.~\cite{Demaine-CO} for a general overview).  Despite these theoretical advances, far less is known about the empirical performance of the techniques developed. The experimental studies that have been carried out into the performance of Cache-Oblivious algorithms (see e.g.~\cite{Tsifakis2004, chatterjee00, brodal02search,ladner02search, dunkel,  Kumar2003, Brodal2007}) have largely focused on internal memory performance in order to test $L1$ and $L2$ cache performance. One notable exception is a recent study where the fastest Cache-Aware and Cache-Oblivious sorting algorithms are also compared in external memory~\cite{Brodal2007}.


Our focus here is on the empirical performance of Cache-Oblivious priority queues and their application to Dijkstra's single source shortest path algorithm for data sizes too large to fit in internal memory. Four Cache-Oblivious priority queues have been developed which we name Arge Heap\cite{ArgeHeap}, Funnel Heap\cite{FunnelHeap}, Bucket Heap\cite{BrodalBucket} and Buffer Heap\cite{Chowdhury}. Although typically these structures have optimal or near optimal asymptotic performance for the operations they support, none so far supports all three of {\sc DecreaseKey}, {\sc Insert} and {\sc DeleteMin} needed for a standard implementation of Dijkstra's algorithm (see Section~\ref{sec:implementation} for a description of some of the modifications required).  The I/O complexity for each priority queue is shown in Figure \ref{Heap-Imp} where we include results for a Cache-Aware tournament tree~\cite{Kumar}, a Cache-Aware priority queue~\cite{Kumar} and Cache-Oblivious tournament trees~\cite{Chowdhury} for completeness.

\begin{table} [h]
   \caption{The I/O complexity of different priority queues}
   \label{Heap-Imp}
   \begin{center}
   \begin{tabular}{|c||c|c|c|c|}
   \hline
               {\bf Priority Queue}  & $Insert$ & $DeleteMin$&
$DecreaseKey$ & $Update$ \\
   \hline
   \hline

               Binary Heap & \multicolumn{4}{|c|}{$O(\log{N})$} \\
   \hline
   \hline
               Cache-Aware Priority Queue &
\multicolumn{2}{|c|}{$O(\frac{1}{B}\log_{\frac{M}{B}}{\frac{N}{B}})$}  & 
- & -\\
   \hline
               Funnel Heap &
\multicolumn{2}{|c|}{$O(\frac{1}{B}\log_{\frac{M}{B}}{\frac{N}{B}})$} & - & -\\
   \hline
               Arge Heap &
\multicolumn{2}{|c|}{$O(\frac{1}{B}\log_{\frac{M}{B}}{\frac{N}{B}})$} & - & -\\
   \hline
               Bucket/Buffer Heap & - &
\multicolumn{1}{|c|}{$O(\frac{1}{B}\log{\frac{N}{B}})$} & - & \multicolumn{1}{|c|}{$O(\frac{1}{B}\log{\frac{N}{B}})$} \\

   \hline
   \hline
               Cache-Aware tournament tree &  - &
\multicolumn{2}{|c|}{$O(\frac{1}{B}\log{\frac{N}{B}})$} & -  \\

   \hline
               Cache-Oblivious tournament tree &  - &
$O(\log{\frac{N}{B}})$ & $O(\frac{1}{B}\log{\frac{N}{B}})$ & -   \\
   \hline
   \end{tabular}
   \end{center}

\end{table}

Figure~\ref{DJ-Imp} gives the corresponding I/O and time complexities for the modified Dijkstra's algorithms evaluated in this study.  Faster asymptotic bounds can be derived in the Cache-Aware model~\cite{Meyer} or if the graphs have bounded weights~\cite{Allulli} or are planar~\cite{Jampala}.
 

\begin{table} [h]
\caption{ The I/O complexities of Dijkstra's Algorithm for the heaps implemented}
\label{DJ-Imp}
\begin{center} 
\begin{tabular}{|c||c|c|c|}
\hline
			& Binary Heap &  Bucket/Buffer Heap   &  Funnel Heap\\
\hline
I/O complexity & $O(E \log{V})$ & $O(V+\frac{E}{B}\log{\frac{V}{B}})$ & $O(V+\frac{E}{B}\log_{\frac{M}{B}}{\frac{V}{B}})$    \\
\hline
\end{tabular}
\end{center}

\end{table}

In this paper we implement Bucket~\cite{BrodalBucket} and Funnel Heap~\cite{FunnelHeap} and compare their performance both to each other and to a standard Binary Heap implementation.  These priority queues are representative of the two main approaches that have been taken. Our preliminary implementation of Buffer Heap for example (not shown here), indicates that its performance tracks that of Bucket Heap closely but is marginally slower in all cases.  We then implement Dijkstra's single source shortest path algorithm using the same priority queues and run a series of tests on both random and real world graph data.   We show that algorithms not explicitly designed for external memory suffer a dramatic performance penalty compared to the Cache-Oblivious algorithms we implement when data is too large to hold in RAM.

\subsection*{Results}

Our main findings are:
 
 \begin{itemize}
 \item For small problem sizes the Binary Heap consistently outperformed the two Cache-Oblivious solutions. This shows that the advantages of optimal multi-level cache usage are outweighed by the constant factor overheads of the more complicated Cache-Oblivious algorithms.
 \item For problem sizes too large to fit in RAM, both the Funnel and Bucket Heap show considerable speedups over Binary Heap on our tests.  For example, using $16$MB of RAM and $1.2$ million elements, Binary Heap took over $4$ hours and spent $>$99\% of its time waiting for I/O requests. Funnel and Bucket Heap by contrast took under $4$ minutes and $10$ minutes respectively.
 \item The performance of Dijkstra's algorithm implemented using the Cache-Oblivious priority queues also showed speedups for large inputs for both synthetic and real world graphs. However, as predicted by the theory these speedups are markedly lower than for the simple priority queue tests due to the cost of accessing the edges in the graph itself. For example, on a graph of $\sim$1 million vertices ($\sim$8 million edges) using $16$MB of RAM for the priority queue and a further $16$MB for the graph, Dijkstra's algorithm implemented with  Funnel Heap was $5$ times faster than using Binary Heap and $20\%$ faster than Bucket Heap.
 \end{itemize}

\section{Implementation}\label{sec:implementation}

All code was written in C++ and compiled using the g++ 4.1.2 compiler with optimisation
level -O3 on GNU$/$Linux distribution Ubuntu $7.10$ (kernel version
$2.6$) with a dual $1.7$ Ghz Intel Xeon
processor PC (only one was used), 1280MB of RAM, 8KB L1 and 256KB L2 cache. The test setup made use of the STXXL Library~\cite{STXXL} version $1.0e$ which is designed to be an STL replacement for processing of large data for experimental testing of external memory algorithms (hence ST\emph{XXL}).The library provides containers and algorithms for large datasets which do not fit in internal memory and handles all swapping of data to and from external storage. As STXXL is designed for Cache-Aware implementations, only a minimal subset of the features available was used with the chosen values of $M$ and $B$ not available to the implemented algorithms.  In order to set up a realistic Cache-Oblivious environment, each algorithm uses one STXXL Vector\footnote{A dynamic array equivalent to the STL vector} for the priority queue and in the case of Dijkstra's algorithm, a further Vector of the same size to store the adjacency list. STXXL Vectors have individual caches which we set to $16$MB and the block size $B$ was set to $4096$ bytes.  The block replacement policy was chosen to be Least Recently Used (LRU). Each machine also has two hard disk
drives, a primary drive containing the Linux boot sector and secondary drive assigned exclusively to the STXXL Library. Both drives perform at $7,200$ RPM with $8.5$ms seek time, $8$MB data buffer with separate parallel ATA 133 interfaces and no secondary cable use. All tests were run in single user mode with the operating system swapping turned off so that STXXL is solely responsible for moving blocks of data in and out of memory. For each test we output the total (wall) time and the I/O wait time as measured by STXXL.

Some further implementation details for the specific tasks carried out follow.

\subsubsection*{Binary Heap} Our Binary Heap implementation is array based with implicit pointers. The {\sc DecreaseKey} operation requires knowledge of the location of the element to be decreased. Maintaining an array of the locations of nodes in the heap requires $O(\log{N})$ I/Os.

\subsubsection*{Funnel Heap}




Funnel Heap was implemented following the description in~\cite{Demaine-CO}.  An important limitation of the Funnel Heap is that it doesn't support the {\sc DecreaseKey} operation. We modified Dijkstra's algorithm to replace all {\sc DecreaseKey} operations with an {\sc Insert} operation instead.  A bit vector is then required to record which vertices have been seen before. This bit vector has size $V$ bits but is required to be kept in internal memory separately from the STXXL Vectors.  The problem is mitigated by the fact that the bitset is small compared to both the adjacency list of the graph and the priority queue data structure that is built. Without the use of an internal memory bit vector the I/O complexity of Dijkstra's algorithm using a Funnel Heap is $O(E+\frac{E}{B}\log_{\frac{M}{B}}{\frac{E}{B}})$.







\subsubsection{Bucket Heap}

Bucket Heap was implemented following the description in~\cite{BrodalBucket}. The Bucket Heap implements an {\sc Update} operation instead of {\sc Insert} and {\sc DecreaseKey} operations. The {\sc Update} operation acts as an {\sc Insert} if the element is not already in the heap and a {\sc DecreaseKey} otherwise. This creates the complication that once a vertex has been removed from the heap and settled it may be re-inserted later by an {\sc Update} operation (acting as an edge relaxation). As we do not want to {\sc DeleteMin} any vertex more than once, this re-insertion must be undone by deleting the element. The problem is to identify which elements are to be deleted. To solve this problem we deploy a technique given by Kumar and Schwabe~\cite{Kumar} for external memory tournament trees. In summary, we allow spurious {\sc Update}s to occur and then delete them before they can be returned by the {\sc DeleteMin}. To identify these spurious updates a second heap is introduced which has an {\sc Update} performed on it for every relaxation of the first heap.  However, some modification of the original method of ~\cite{BrodalBucket} is required to be able to handle the case where the two heaps return elements with identical keys. We therefore process the elements from the second heap twice, once before elements from the main heap and then again afterwards. The modification leaves the asymptotic I/O complexity of Dijkstra's algorithm unchanged.

\section{Results and Analysis} \label{sec:results}

In this Section we present the main experimental results. A single STXXL Vector was used for each priority queue test with associated cache size $M$ set to $16$MB. In the tests of Dijkstra's algorithm, an additional STXXL Vector with associated cache size $16$MB  was used to store the input graph. A further set of tests was also carried out to test the effect of varying the cache size.  
\subsection*{Priority queue tests} \label{DPQ}

We tested the performance of the Binary, Funnel and Bucket Heap  by performing a simple sequence of {\sc Insert} and {\sc DeleteMin} operations: 

	\begin{enumerate}
	\item
			{\sc Insert} $N$ elements with randomly chosen priorities.
	\item
			Perform $\lfloor \frac{N}{2} \rfloor$ {\sc DeleteMin}s.
	\item
			{\sc Insert} $\lfloor \frac{N}{2} \rfloor$ elements with randomly chosen priorities.
	\item
		Perform $N$ {\sc DeleteMin}s, leaving the heap empty at the end.
	\end{enumerate}
	
Each test was terminated automatically if the runtime exceeded 6 hours and results quoted are averages over three runs. We also ran a second set of tests over the range during which Binary Heap began swapping which were not repeated due to the length of time they took to run. 

\begin{figure}[h]
\centering
\includegraphics[width=\textwidth]{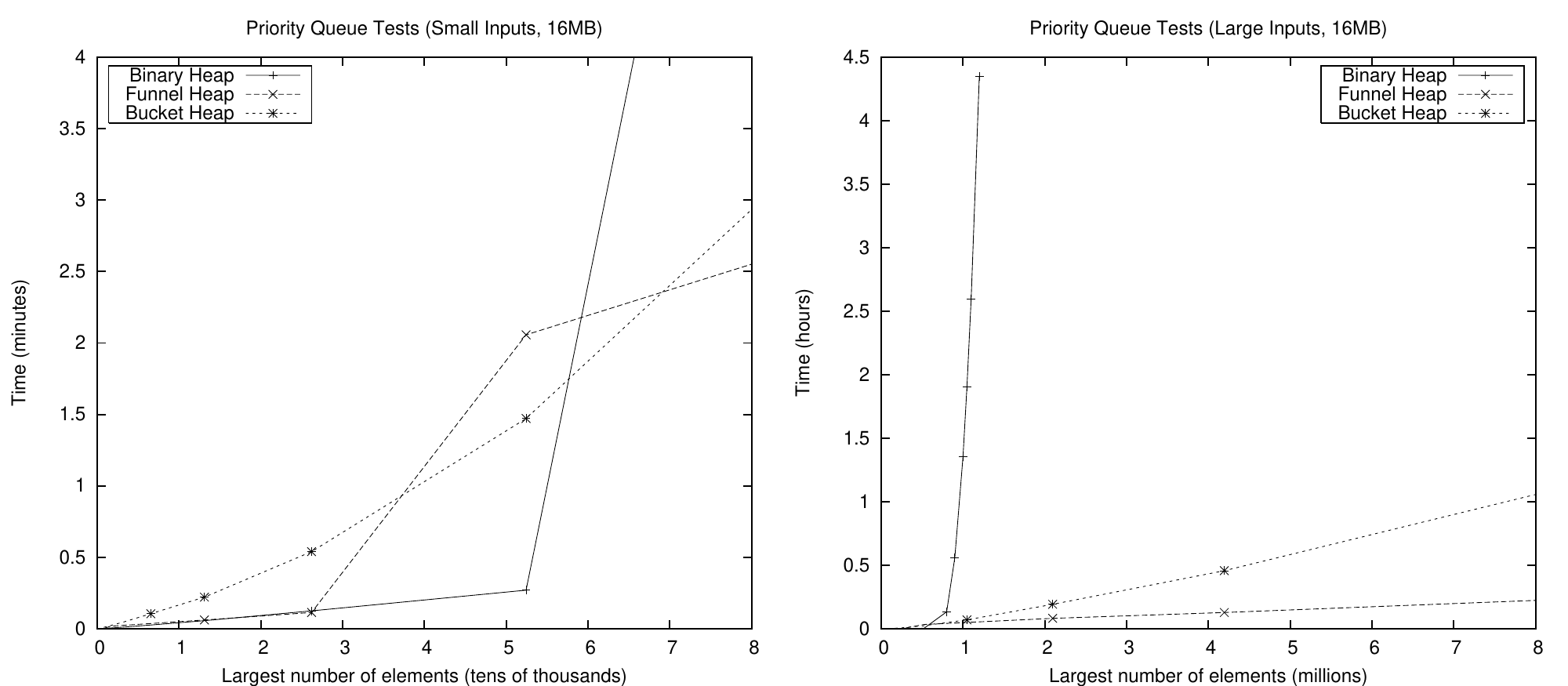}
\caption{Total time taken for Priority Queue tests}
\label{FIG-PQ}
\end{figure}

Figure~\ref{FIG-PQ} and Table~\ref{TAB-PQ} show the results for increasing numbers of elements. When the input size is small enough that Binary Heap fits inside memory its performance is consistently superior to the two external memory heaps. Due to the much higher space requirements of Funnel and Bucket Heap both also started swapping earlier than Binary Heap. As an example, for $524288$ elements Binary Heap spends $<5\%$ of the total time waiting for I/O requests while Funnel Heap and Bucket Heap spend $\sim$73\% and $\sim$32\% respectively. After $\sim$0.7 million elements Binary Heap starts to swap and slows down dramatically.

Funnel and Bucket Heap continue to perform well even once all three structures are swapping heavily.  Funnel Heap completes on $\sim$33 million elements in less time than Binary Heap on $1$ million elements and approximately the same time as Bucket Heap on $\sim$8 million elements.  The superior performance of Funnel Heap  is likely to be for a number of reasons. Not only does it have an $O(\log{\frac{M}{B}})$ factor lower asymptotic complexity but it is also a considerably less complicated structure than the Bucket Heap. Another advantage the Funnel Heap has over Bucket Heap is the use of an extra $V$ internal bits which are not swapped out (see Section~\ref{sec:implementation}).  We also note that the percentage I/O wait time for Funnel Heap has considerable fluctuations. This is because the heap grows in increasingly large jumps as each additional funnel is added. That is, the addition of a single element may require the construction of an entire funnel. It is possible that this fluctuation could be removed by part building/expanding the funnels only when they are needed. 
 
\begin{table}
\centering
\caption{Priority queue tests with 16MB internal memory}
\begin{tabular}{l|cc|cc|cc}
\hline
\noalign{\smallskip}
  \multirow{2}{*}{Size} &\multicolumn{2}{|c|}{Binary Heap}  &  \multicolumn{2}{|c|}{Funnel Heap} &  \multicolumn{2}{|c}{Bucket Heap} \\
  & Time (s)  &  I/O wait (\%)  & Time (s)  &  I/O wait  (\%) & Time  (s) &  I/O wait  (\%) \\
\noalign{\smallskip}
\hline
\noalign{\smallskip}
65536 & 2 & - & 2 & 28.9\% & 6 & 9.6\% \\
131072 & 3 & - & 4 & 17.0\% & 13 & 6.2\% \\
262144 & 8 & - & 7 & 9.5\% & 32 & 15.2\% \\
524288 & 16 & 4.4\% & 123 & 73.3\% & 88 & 32.2\% \\
1048576 & 6850 & 99.1\% & 180 & 70.9\% & 255 & 49.1\% \\
2097152 & $>$6 hrs & - & 299 & 66.9\% & 694 & 59.5\% \\
4194304 & - & - & 463 & 63.7\% & 1649 & 63.5\% \\
8388608 & - & - & 843 & 62.7\% & 4028 & 68.0\% \\
16777216 &  - & - & 1756 & 64.3\% & 9792 & 71.5\% \\
33554432 & - & - & 4057 & 69.5\% & $>$6 hrs & - \\
\noalign{\smallskip}
\hline
\end{tabular} \label{TAB-PQ}
\end{table}

\begin{figure}[h]

\centering
\includegraphics[width=\textwidth]{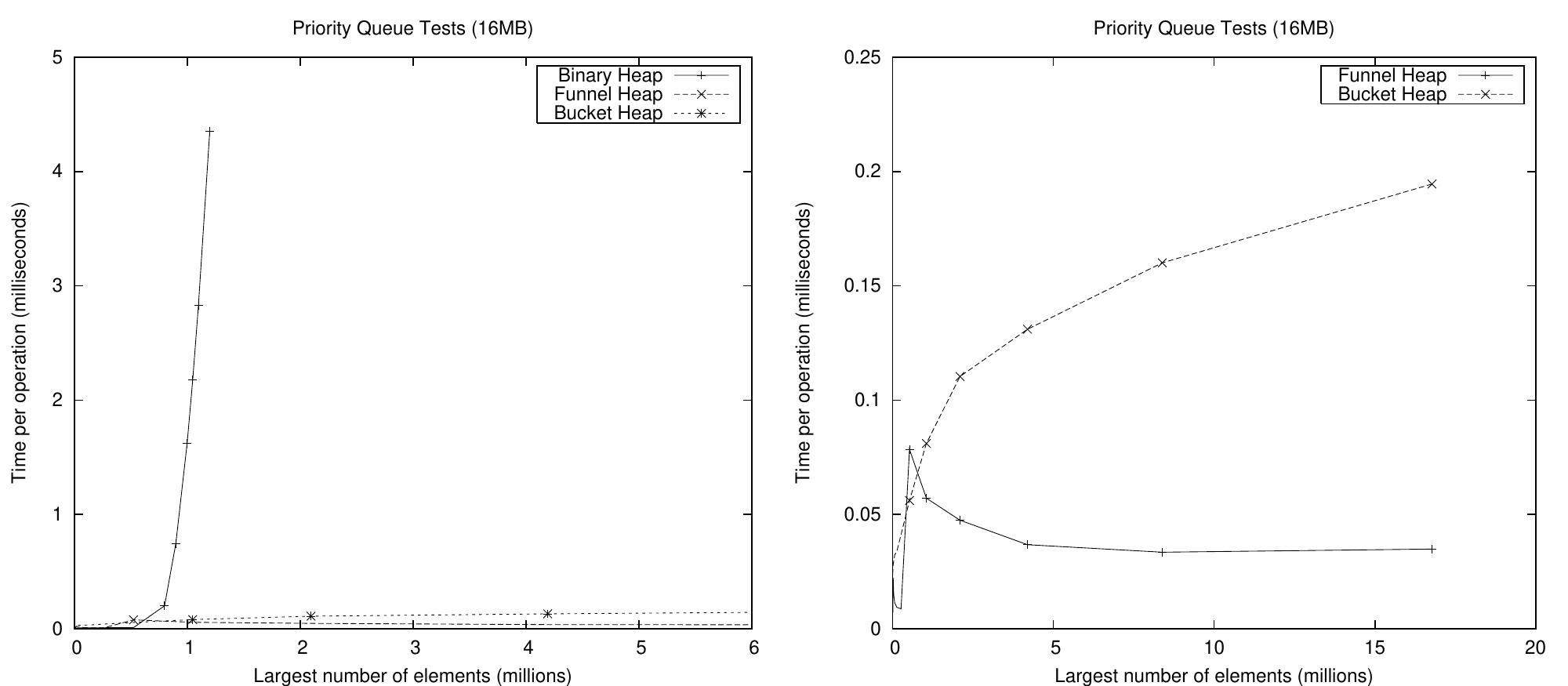}
\caption{Time taken per element for Priority Queue tests}
\label{FIG-PQPER}
\end{figure}

Figure \ref{FIG-PQPER} gives the time taken per operation for each of the heaps. The time per operation for Binary Heap rapidly exceeds 4ms while it remains below 0.2ms for both Bucket and Funnel Heap even past $16$ milion elements. Funnel Heap's time per operation is strongly affected by how recently a new funnel has been created, particularly for small input however its overall superior performance is again clear.

The same tests carried out with $M$ set to $128$MB showed almost exactly equivalent but suitably scaled results. The details are omitted for reasons of space. 

%
%



	


\subsubsection*{Shortest path tests} The graphs were generated according to the Erd\"{o}s-R\'{e}nyi ($G(n,p)$) model. In this model the structure of a graph is generated based on two parameters, the number of vertices, $n$ and a probability, $p$, of each edge existing. We used integer weights and $p=\frac{16}{V-1}$ giving an expected $E=8V$ edges. The graphs were undirected and all results are averaged across three test runs.

\begin{figure}[h]
\centering
\includegraphics[width=\textwidth]{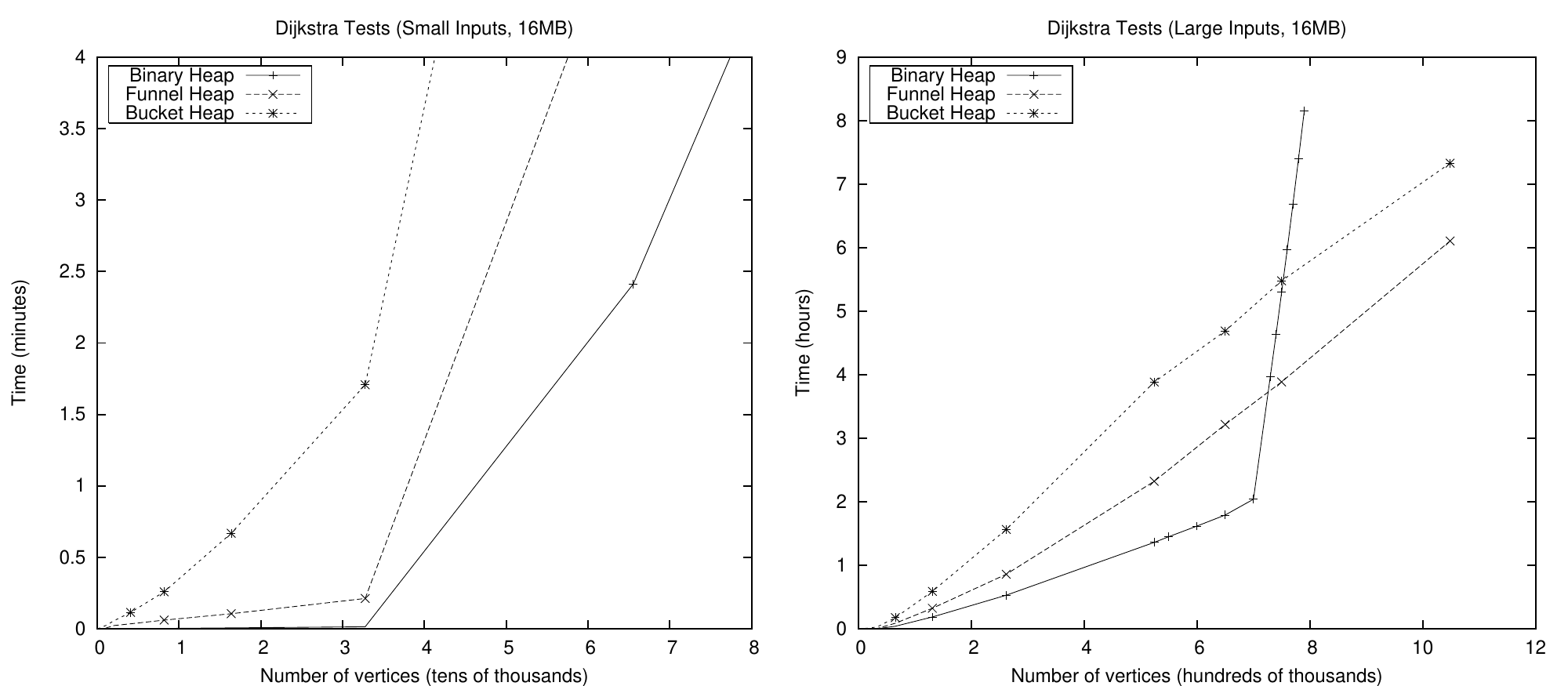}
\caption{Total time taken in Dijkstra's Algorithm tests on random graphs}
\label{FIG-DJ}
\end{figure}

Figure~\ref{FIG-DJ} (right) shows the performance of Dijkstra's algorithm run to completion on random graphs with the start nodes also chosen at random. As before Binary Heap performs well for small graphs but Table~\ref{TAB-DJ-RAN} shows that as the number of vertices increases from $\sim$0.75 to $\sim$1 million vertices Binary Heap's running time increases by a factor of $\sim$5.8. Here Funnel Heap's performance is much closer to Bucket Heap's than in the previous tests. The modifications made to Dijkstra's algorithm to account for Funnel Heap's lack of a {\sc DecreaseKey} operation mean that the heap contains $O(E)$ elements not $O(V)$ elements. While this does not affect the asymptotic complexity, it is likely to be the main contributor to the decreased separation between the performace of Bucket and Funnel Heap. Figure~\ref{FIG-DJ} (left) shows the performance for graphs small enough that the priority queues fit completely in RAM.  It can clearly be seen that the structures containing $O(E)$ elements start to swap heavily at $\sim$35 thousand vertices.

\begin{table}
\centering
\caption{Selected Dijkstra's Algorithm tests on random graphs with 16MB internal memory}
\begin{tabular}{l|cc|cc|cc}
\hline
\noalign{\smallskip}
  \multirow{2}{*}{Vertices} &\multicolumn{2}{|c|}{Binary Heap}  &  \multicolumn{2}{|c|}{Funnel Heap} &  \multicolumn{2}{|c}{Bucket Heap} \\
  & Time (s)  &  I/O wait (\%)  & Time (s)  &  I/O wait  (\%) & Time  (s) &  I/O wait  (\%) \\
\noalign{\smallskip}
\hline
\noalign{\smallskip}
65536 & 145 &  98.2\% & 313 & 84.6\% & 631 & 63.2\% \\
131072 & 672 & 98.9\% & 1167 & 92.6\% & 2124 & 78.6\% \\
262144 & 1908 &  99.1\% & 3095 & 94.6\% & 5628 & 83.1\% \\
524288 & 4895 & 99.1\% & 8361 & 96.1\% & 13987 & 84.6\% \\
750000 &   19100 & 99.4\% & 13995 & 96.6\% & 19725 & 83.8\% \\
1048576 & 111061 & 99.7\% & 21984 & 97.0\% & 26389 & 83.4\% \\ 
\hline
\noalign{\smallskip}
\end{tabular} \label{TAB-DJ-RAN}
\end{table}

\subsubsection*{Real world graphs}

\begin{figure}[h]
\centering
\begin{tabular}{cc}

\begin{minipage}{6cm}
\centering
\includegraphics[width=6cm]{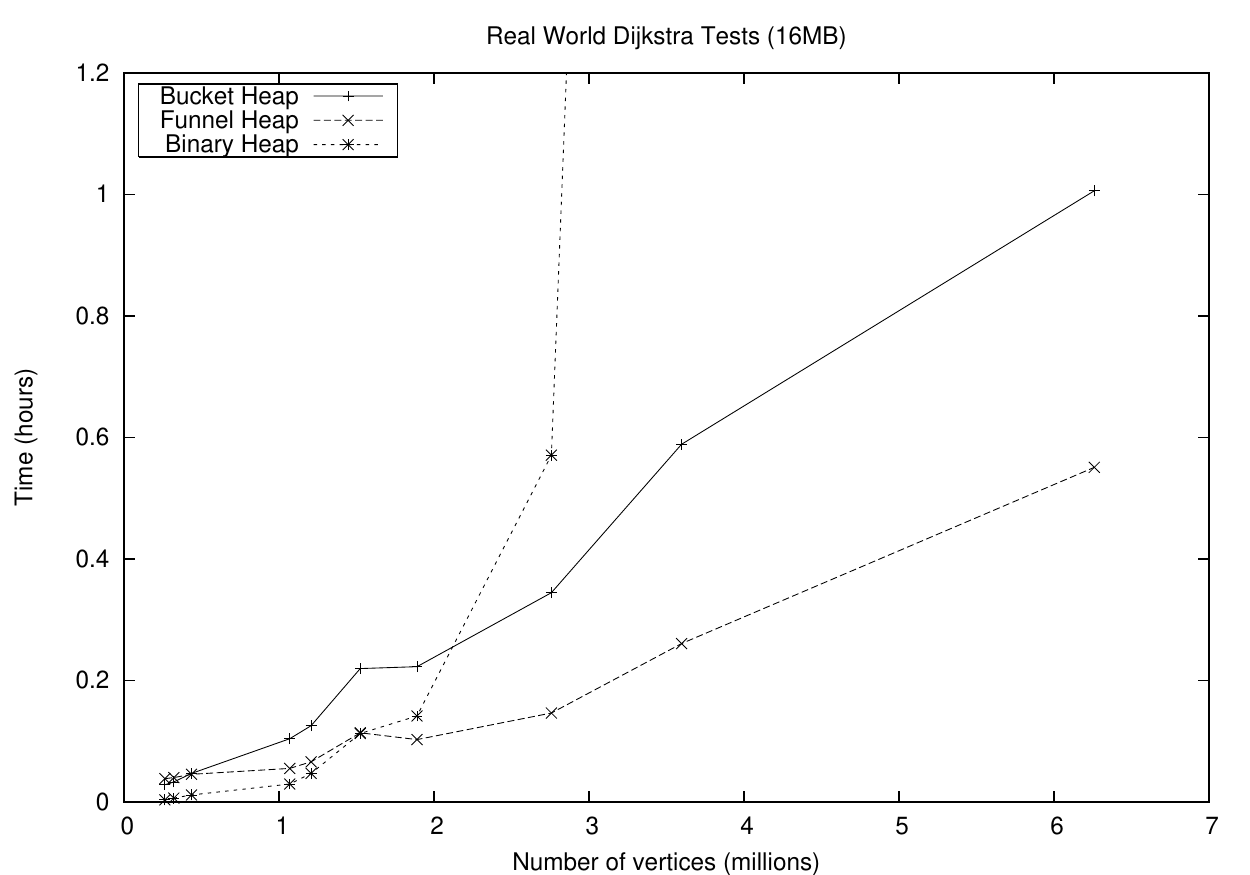}
\end{minipage}

&

\begin{minipage}{6cm}
\centering
\includegraphics[width=6cm]{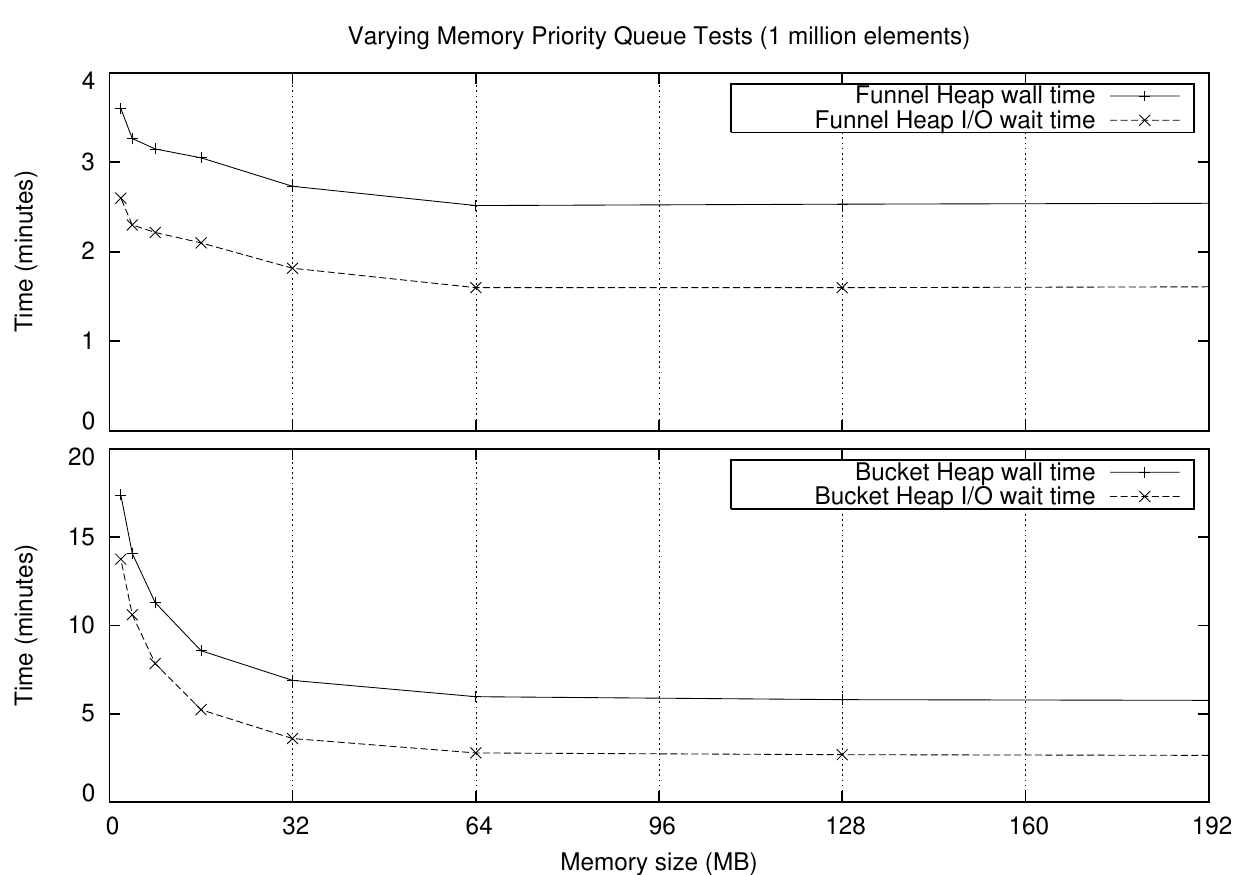}
\end{minipage}

\end{tabular}

\caption{Total time taken for Dijkstra's algorithm on real world graphs (left) and the effects of varying memory size on the simple priority queue tests (right)} \label{FIG-DJ+MEM}
\end{figure}

In addition to randomly generated graphs, the algorithms were run on real world graphs\footnote{The DIMACS SSSP challenge graphs are undirected versions of the major road networks of the United States of America} from the DIMACS shortest path challenge\cite{DIMACSP}.  These graphs are almost planar and as a result sparse with $E<3V$ in all cases. Figure \ref{FIG-DJ+MEM} shows the performance on these graphs. Funnel Heap performs better on the real world graphs than on the random graphs due to this increased sparseness.  This reduces the overhead caused by having to store $O(E)$ elements in its heap.


\begin{table}
\centering
\caption{Dijkstra's Algorithm tests on real world graphs with 16MB internal memory}
\begin{tabular}{l|cc|cc|cc}
\hline
\noalign{\smallskip}
  \multirow{2}{*}{Vertices}  & \multicolumn{2}{|c|}{Binary Heap}  &  \multicolumn{2}{|c|}{Funnel Heap} &  \multicolumn{2}{|c}{Bucket Heap} \\
   & Time (s) &  I/O wait (\%)  & Time (s)  &  I/O wait  (\%) & Time (s) &  I/O wait (\%)\\
\noalign{\smallskip}
\hline
\noalign{\smallskip}
264346  & 14 & 66.2\% & 138 & 78.6\% & 102 & 10.0\% \\
321270 & 24 & 74.8\% & 146 & 79.9\% & 120 & 17.2\% \\
435666  & 42 & 78.6\% & 164 & 80.4\% & 169 & 22.4\% \\
1070376  & 107 & 76.4\% & 200 & 74.1\% & 376 & 13.8\% \\
1207945  & 169 & 82.5\% & 238 & 75.9\% & 452 & 20.7\% \\
1524453  & 406 & 90.0\% & 411 & 79.1\% & 791 & 29.7\% \\
1890815  & 510 & 90.1\% & 371 & 76.7\% & 802 & 25.8\% \\
2758119  & 2055 & 95.8\% & 528 & 77.1\% & 1240 & 27.5\% \\
3598623  & $>$6 hrs & - & 939 & 81.9\% & 2120 & 30.2\% \\
6262104  & - & - & 1983 & 85.6\% & 3622 & 37.8\% \\
\noalign{\smallskip}
\hline
\end{tabular}
\end{table}


\subsubsection*{Varying internal memory size} We investigated the effect of varying the internal memory size on the Cache-Oblivious priority queues with 1 million elements.  This size was chosen as it is small enough to be reasonably fast to compute but large enough that both Bucket and Funnel Heap are swapping heavily for all but the largest memory sizes. With 1 million elements, Bucket and Funnel Heap use $\sim$125MB and $\sim$160MB space respectively. We ran tests for memory size of up to $1024$MBs (repeated 3 times). Figure \ref{FIG-DJ+MEM} shows that even once the structures fit completely into memory some I/O is still reported by STXXL.  This is due to the set up costs of creating the STXXL Vectors.  The most remarkable aspect of the results is that Funnel Heap with $2$MB of memory outperforms Bucket Heap with $1024$MB of memory. It is also of interest that Bucket Heap appears to be affected far more by varying memory than Funnel Heap. This is a property of the data structures which is not fully captured by their asymptotic I/O complexity.

\subsubsection*{Acknowledgment} The authors would like to thank Ashley Montanaro for helpful comments on the final draft.

\bibliography{myrefs90}        
\end{document}